\documentclass[12pt,a4paper]{article}

\usepackage{graphicx}
\usepackage{cite}
\usepackage{bm}
\usepackage{color}

\begin{document}
\begin{center}
\Large{\bf Phase diagram of  Ca$_{1-x}$Ce$_{x}$MnO$_{3}$ thin films studied by x-ray magnetic  circular dichroism \\}

\vspace{12pt}

\normalsize{T. Harano$^1$, G. Shibata$^1$, K. Yoshimatsu$^1$, K. Ishigami$^2$, V.K.Verma$^1$, Y. Takahashi$^1$, \\
T. Kadono$^1$, T. Yoshida$^1$, A. Fujimori$^1$, T. Koide$^3$, F.-H. Chang$^4$, H.-J. Lin$^4$, D.-J. Huang$^4$, \\
C.-T. Chen$^4$, P.-H. Xiang$^{5,\ 6}$, H. Yamada$^5$, and A. Sawa$^5$ \\

}

\vspace{12pt}

{\it 
$^1$Department of Physics, University of Tokyo, Bunkyo-ku, Tokyo 113-0033, Japan \\
$^2$Department of Complexity Science and Engineering, University of Tokyo, Kashiwa-shi, Chiba 277-8561, Japan \\
$^3$High Energy Accelerator Research Organization (KEK) Photon Factory, Tsukuba, Ibaraki 305-0016, Japan \\
$^4$National Synchrotron Radiation Research Center (NSRRC), Hsinchu 300, Taiwan \\
$^5$National Institute of Advanced Industrial Science and Technology (AIST), Tsukuba, Ibaraki 305-8562, Japan \\
$^6$Key Laboratory of Polar Materials and Devices, Ministry of Education, East China Normal University, Shanghai 200241, China. \\
}

\end{center}

\section*{Abstract}
In the perovskite-type Ca$_{1-x}$Ce$_{x}$MnO$_{3}$ (CCMO), one can control the transport and magnetic properties through varying Ce content. In the case of thin films, the properties can also be controlled by epitaxial strain from the substrate through changing it such as YAlO$_{3}$ (YAO), NdAlO$_{3}$ (NAO), and LaSrAlO$_{4}$ (LSAO). However, one cannot measure the magnetization of thin films on NAO substrates by conventional magnetization measurements because of the strong paramagnetic signals from the Nd$^{3+}$ ions. In order to eliminate the influence of Nd$^{3+}$ and to identify magnetic phases of the CCMO thin films, we have performed element-selective X-ray magnetic circular dichroism (XMCD) measurements of the Mn 2{\it p} core level. By studying the anisotropy of the XMCD intensity, we could unambiguously determine the magnetic phase diagram of the CCMO thin films. 

\vspace{36pt}

The perovskite-type manganites, which have the chemical formula of {\it R}$_{1-x}${\it A}$_{x}$MnO$_{3}$ ({\it R}: rare-earth, {\it A}: alkali-earth), have attracted considerable interest because one can control the spin, orbital, and charge orderings by changing {\it R}, {\it A}, and {\it x} \cite{Review_manganitebulk}. The manganites have also attracted attention as materials showing giant magnetroresistance. In the case of thin films, one can also control their physical properties by changing the in-plane lattice constants through epitaxial strain from the substrate \cite{KonishiJPSJ}.

\begin{figure}[hbt]
\begin{center}
\includegraphics[width=12cm]{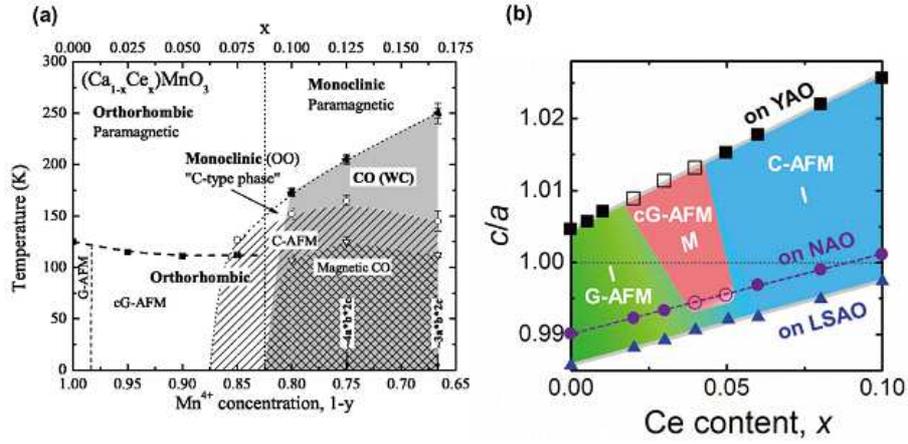}
\caption{(a) Temperature-composition phase diagram of bulk Ca$_{1-x}$Ce$_{x}$MnO$_{3}$ (CCMO) \cite{PD_CCMObulk}. (b) Lattice strain ({\it c/a})-composition ({\it x}) phase diagram at low temperatures of CCMO thin films grown on {\it c}-axis-oriented YAlO$_{3}$ (YAO), NdAlO$_{3}$ (NAO), and LaSrAlO$_{4}$ (LSAO) substrates \cite{PD_CCMOfilm, mx}. CO: Charge ordered, OO: Orbital ordered, WC: Wigner crystal, G-AFM: G-type antiferromagnetic, C-AFM: C-type antiferromagnetic, cG-AFM: Canted G-type antiferromagnetic, M: metallic, I: Insulating.}
\label{PD}
\end{center}
\end{figure}

While extensively studies have been done for hole-doped manganites such as La$_{1-x}$Sr$_{x}$MnO$_{3}$ (LSMO), electron-doped manganites such as Ca$_{1-x}$Ce$_{x}$MnO$_{3}$ (CCMO) have also been studied for their interesting interplay between spin, charge, and orbital ordering. In bulk CCMO, one can control the transport properties and the ordered spin structures, indicated by the temperature-composition phase diagram as shown in Fig. \ref{PD} (a) \cite{PD_CCMObulk}. In CaMnO$_3$, the G-type antiferromagnetic (G-AFM) state is stable with the spin moments directed along the {\it c}-axis of the orthorombically distorted perovskite structure. With Ce doping, the spins of the G-AFM state start to cant at around {\it x} = 0.012 towards the {\it a}-axis \cite{Ohnishi}, and the canted G-AFM (cG-AFM) state is realized. Moreover, for larger Ce content {\it x} $\geq$ 0.0625, the C-type AFM phase, where the spins are aligned within the {\it ab}-plane, appears. There are other phases in the region {\it x} $\geq$ 0.0865 such as charge-ordered (CO), Wigner-crystal-type ordered (WC), and orbital-ordered (OO) phases. As for thin films of CCMO, Xiang {\it et al.} \cite{PD_CCMOfilm, mx} have reported that the transport properties and magnetic ordering can also be controlled by changing the substrate such as {\it c}-axis-oriented YAlO$_{3}$ (YAO), NdAlO$_{3}$ (NAO), and LaSrAlO$_{4}$ (LSAO). The phase diagram of CCMO thin films for different substrates and Ce content {\it x} has been determined using transport and magnetization measurements shown in Fig. \ref{PD} (b). However, one cannot measure the magnetization of CCMO thin films grown on NAO substrates because of the strong paramagnetic signals from the concentrated Nd$^{3+}$ ions in the substrate.


X-ray magnetic circular dichroism (XMCD) is a powerful technique to obtain the element-specific magnetic information of complex systems. Therefore, one can extract the magnetization of Mn atoms in samples grown on NAO substrates. In the present work, in order to identify the different magnetic phases of CCMO thin films grown on the various substrates, we have measured the magnetic field direction dependence of the Mn XMCD intensity for various CCMO thin films.


In the pseudocubic notation, {\it c}-axis-oriented and single-crystalline thin films of CCMO (0 $\leq$ {\it x} $\leq$ 0.1) with thickness of 40 nm were fabricated on YAO, NAO, and LSAO substrates by the pulsed-laser deposition method. All the samples are listed in Table \ref{table} together with their easy magnetization axes \cite{mx} and expected spin structures and directions \cite{PD_CCMObulk, Ohnishi, mx}. As mentioned above, however, the information about the films grown on NAO substrates  has been limited to that inferred from the transport data \cite{PD_CCMOfilm} and the information about the magnetic properties was first obtained in the present work, as described below. The detailed deposition conditions were reported elsewhere \cite{PD_CCMOfilm}. After cooling down to room temperature, the samples were capped with amorphous LaAlO$_{3}$ layers (1-2 nm thickness) to prevent contamination. Four-circle X-ray diffraction measurements were performed to evaluate the thickness, the lattice parameters, and the crystal structure. The reciprocal space maps of the CCMO films revealed that the films were coherently grown on the substrates. The {\it x} dependence of the strain states of the CCMO films has been previously reported \cite{PD_CCMOfilm, mx}. For all values of {\it x} examined in this study, the films on YAO are subjected to a compressive strain, whereas the films on LSAO are subjected to a tensile strain. In the case of NAO substrates, the films with {\it x} = 0-0.06 are subjected to a tensile strain, whereas the films with {\it x} = 0.08-0.1 are practically strain-free.
The CCMO film with {\it x} = 0.04 grown on the YAO substrate has an in-plane compressive strain of -0.8 $\%$, whereas those grown on NAO and LSAO have in-plane tensile strain of +0.3 $\%$ and +0.5 $\%$, respectively \cite{PD_CCMOfilm, mx}. 
The XMCD measurements were performed at BL-11A of National Synchrotron Radiation Research Center (NSRRC), Taiwan, and BL-16A of Photon Factory, KEK, Japan. In both measurements, we employed the total electron yield (TEY) method. All the measurements were performed at a temperature of {\it T} = 20 K in a magnetic field of {\it H} = 1 T. At NSRRC, we applied the magnetic field at $\theta$ = 0$^{\circ}$ and 90$^{\circ}$, where $\theta$ is defined as the angle between magnetic field and the sample surface normal. At KEK, we applied the magnetic field at $\theta$ = 0$^{\circ}$, 60$^{\circ}$, and 75$^{\circ}$. 

\begin{figure}[hbt]
\begin{center}
\includegraphics[width=15cm]{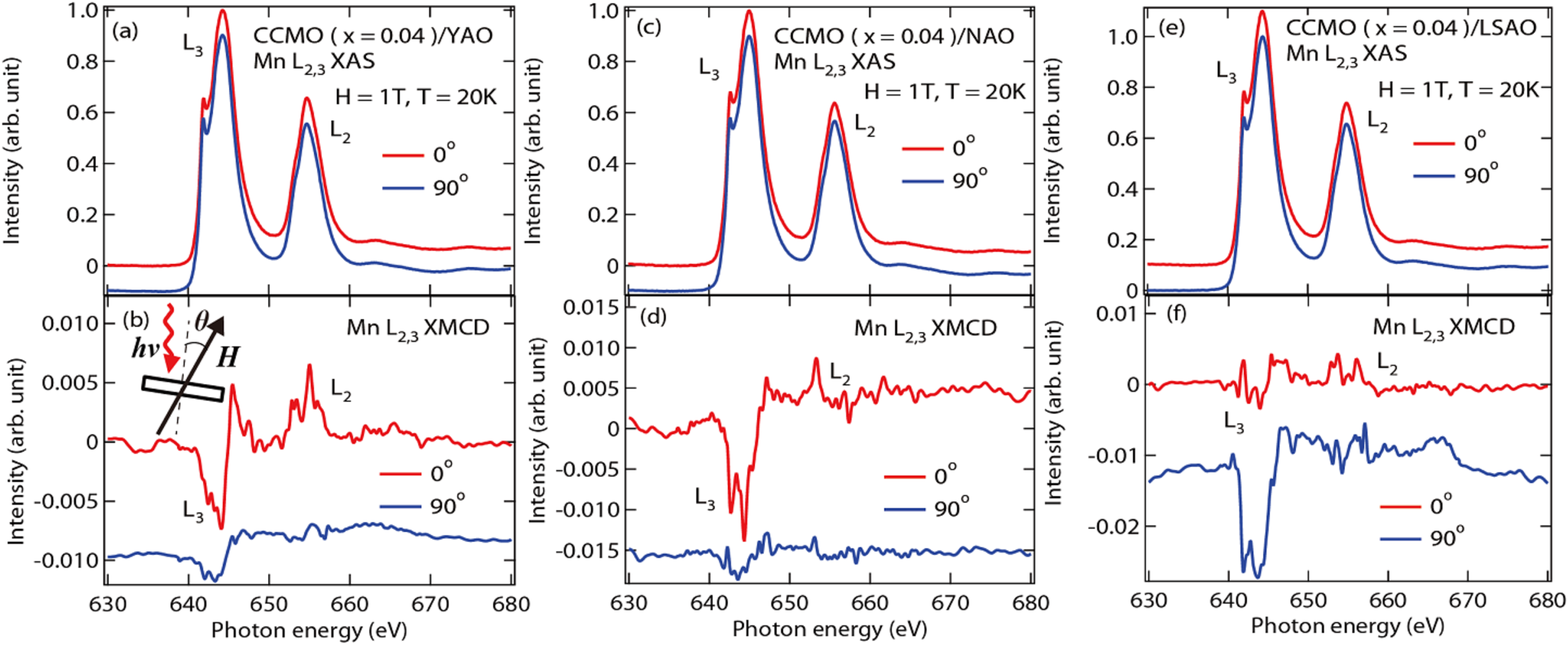}
\caption{Mn 2{\it p}-3{\it d} XAS and XMCD spectra of CCMO {\it x} = 0.04 thin films grown on different  {\it c}-axis-oriented YAO [(a) and (b)], NAO [(c) and (d)], and LSAO [(e) and (f)] substrates. Angle $\theta$ between the magnetic field and the film surface normal was 0$^{\circ}$ or 90$^{\circ}$, as shown in the inset.}
\label{Subdep}
\end{center}
\end{figure}

We have measured two sets of samples. The first set of samples are CCMO thin films with fixed {\it x} ({\it x} = 0.04) grown on the three different substrates, i.e, {\it c}-axis-oriented YAO, NAO, and LSAO. Figure \ref{Subdep} shows the Mn 2{\it p} $\to$ 3{\it d} XAS and XMCD spectra of the samples. Each spectrum consists of the {\it L}$_3$ peak (Mn 2{\it p}$_{3/2}$ $\to$ 3{\it d} absorption peak) and the {\it L}$_{2}$ peak (Mn 2{\it p}$_{1/2}$ $\to$ 3{\it d} absorption peak). The overall spectral line shapes are consistent with the previous XMCD studies \cite{Yanagida, Takamura}. For the LSAO substrate, the XMCD signal is stronger for $\theta$ = 90$^{\circ}$ than for $\theta$ = 0$^{\circ}$. On the other hand, for the YAO and NAO substrates, the {\it L}$_3$ XMCD peak is weaker for $\theta$ = 90$^{\circ}$ than for $\theta$ = 0$^{\circ}$. Xiang {\it et al.} \cite{PD_CCMOfilm} have measured the  magnetization of CCMO thin films by applying magnetic field  perpendicular and parallel to the film surface, and concluded that the easy magnetization axis is perpendicular to the plane for the G-AFM and cG-AFM phases and parallel to the plane for the C-AFM phase. The present XMCD results for the YAO and LSAO substrates are, therefore, consistent with the cG-AFM and G-AFM phases, respectively, and the XMCD results for the NAO substrate is consistent with the cG-AFM phase. 

\begin{figure}[hbt]
\begin{center}
\includegraphics[width=15cm]{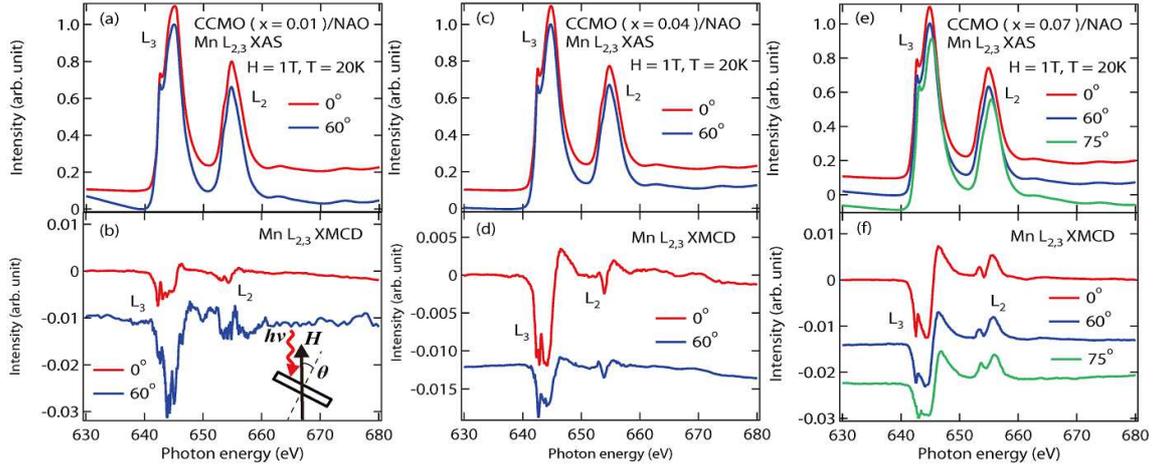}
\caption{Mn 2{\it p}-3{\it d} XAS and XMCD spectra of CCMO thin films with various Ce concentrations grown on {\it c}-axis-oriented NAO substrates: {\it x} = 0.01 [(a) and (b)], 0.04 [(c) and (d)], and 0.07 [(e) and (f)]. Angle $\theta$ between the magnetic field and the film surface normal was 0$^{\circ}$, 60$^{\circ}$, or 75$^{\circ}$, as shown in the inset.}
\label{Cedep}
\end{center}
\end{figure}

The other set of samples we have studied are CCMO thin films with various {\it x}'s ({\it x} = 0.01, 0.04 and 0.07) grown on NAO substrates. Figure \ref{Cedep} shows the Mn 2{\it p} $\to$ 3{\it d} XAS and XMCD spectra of these samples. For {\it x} = 0.01, the XMCD intensity for $\theta$ = 60$^{\circ}$ is stronger than that for $\theta$ = 0$^{\circ}$, consistent with the magnetic anisotropy of the G-AFM phase \cite{PD_CCMOfilm}. On the other hand, for {\it x} = 0.04 and 0.07, the XMCD intensity is stronger for $\theta$ = 0$^{\circ}$ than for $\theta$ = 60$^{\circ}$ and/or 75$^{\circ}$, consistent with the cG-AFM, and C-AFM phases \cite{PD_CCMOfilm}, respectively.


The easy magnetization axes and the inferred spin structures for the CCMO thin films grown on NAO substrates, as well as those grown on YAO and LSAO substrates, are summarized in Table \ref{table}. The directions of the ordered spin moments are considered to be perpendicular to the easy magnetization axes for the G- and C-AFM. The present approach of using XMCD will be useful for various magnetic oxide thin films grown on substrates containing paramagnetic rare-earth ions such as NdGaO$_3$, GdScO$_3$, and DyScO$_3$. 
These substrates have also been used to study the effects of epitaxial strains on the physical properties of magnetic thin films such as La$_{1-x}$Sr$_x$MnO$_3$\cite{Dho,Vailionis}, SrRuO$_3$\cite{Vailionis,Kan}, and BiFeO$_3$. So far, the effects of strain on the structural properties of SrRuO$_3$ and BiFeO$_3$\cite{Johann} thin films have been studied using substrates of NdGaO$_3$, GdScO$_3$, DyScO$_3$, etc., but their magnetic properties have not been studied yet.

In conclusion, we have studied the magnetic anisotropy of CCMO thin films grown on various substrates by measuring the Mn 2{\it p} $\to$ 3{\it d} XMCD intensity for different magnetic field-directions. In particular, for CCMO thin films grown on NAO substrates, whose strong Nd$^{3+}$ paramagnetic signals prohibit conventional magnetization measurements, magnetic anisotropy was the same between {\it x} = 0.04  (cG-type AFM) and {\it x} = 0.07 (C-type AFM) but was different from {\it x}=0.01 (G-type AFM), confirming the phase diagram proposed by Xiang {\it et al.} on the basis of transport \cite{PD_CCMOfilm} and magnetization measurements \cite{mx}. 


We thank H. Ohnishi for useful discussion. Technical advice and support by K. Amemiya and M. Sakamaki are gratefully acknowledged. This work was supported by a Grant-in-Aid for Scientific Research from JSPS (S22224005) and the ``Quantum Beam Technology Development Program" from JST.
The experiment was done under the approval of the Photon Factory Program Advisory Committee (proposal nos. 2012G667 and 2010S2-001). The experiment was done under the approval of the National Synchrotron Radiation Research Center (proposal no. 2011-3-106-2).

\clearpage

\renewcommand{\thetable}{\Roman{table}}

\begin{table}[t]
\begin{center}
\caption{Magnetic properties of Ca$_{1-x}$Ce$_{x}$MnO$_{3}$ (CCMO) thin films grown on {\it c}-axis-oriented substrates of YAlO$_{3}$ (YAO), NdAlO$_{3}$ (NAO) and LaSrAlO$_{4}$ (LSAO) \cite{mx} and bulk Ca$_{1-x}$Ce$_{x}$MnO$_{3}$ crystals \cite{PD_CCMObulk, Ohnishi}. The spin structures for the films grown on NAO substrates are revealed from the easy magnetization directions measured in the present work. ``Out": Out of plane, ``In": In-plane, ``Canted": Canted from in-plane to out-of-plane, G: G-type antiferromagnetic, C: C-type antiferromagnetic, cG: Canted G-type antiferromagnetic. The {\it a}, {\it b}, and {\it c} axes of bulk CCMO are defined for the orthorhomic ({\it Pbmn}) phase.  }
\vspace{1cm}
\begin{tabular}{lcccccccccc}
\hline
 & \multicolumn{7}{c}{Film} & \multicolumn{3}{c}{Bulk} \\
Substrate & YAO & \multicolumn{5}{c}{NAO} & LSAO & \multicolumn{3}{c}{} \\ 
{\it x} & 0.04 & & 0.01 & 0.04 & 0.07 & & 0.04 & 0.01 & 0.04 & 0.07 \\ \hline
Easy magnetization axis & out & & in & out & out & & in & - & - & - \\ 
Spin structure & cG & & G & cG & C & & G & G & cG & C \\ 
Direction of spin moment & Canted & & In/Out & Canted & In & & In/Out & {\it c} & {\it c} $\to$ {\it a} & {\it ab} \\ \hline
\label{table}
\end{tabular}
\end{center}
\end{table} 

\clearpage

\clearpage

\clearpage

\clearpage

\graphicspath{{CCMO/fig/}}

\end{document}